\documentclass[conference]{IEEEtran}
\IEEEoverridecommandlockouts
\usepackage{cite}
\usepackage{amsmath,amssymb,amsfonts}
\usepackage{algorithmic}
\usepackage{graphicx}
\usepackage{textcomp}
\usepackage{xcolor}
\def\BibTeX{{\rm B\kern-.05em{\sc i\kern-.025em b}\kern-.08em
    T\kern-.1667em\lower.7ex\hbox{E}\kern-.125emX}}
\newcommand{\norm}[1]{\left\lVert#1\right\rVert}

\DeclareMathOperator{\sinc}{sinc} 
\DeclareMathOperator{\rect}{rect}

\begin{document}

\title{Exploring the Interdependencies Between Transmit Waveform Ambiguity Function Shape and Off-Axis Bearing Estimation
\thanks{This research was supported by ONR grant N0001421WX00843}
}

\author{\IEEEauthorblockN{Matthew D. Tidwell}
\IEEEauthorblockA{\textit{Sensors and Sonar Systems Department} \\
\textit{Naval Undersea Warfare Center}\\
Newport, RI, USA \\
matthew.d.tidwell@navy.mil}
\and
\IEEEauthorblockN{David A. Hague}
\IEEEauthorblockA{\textit{Sensors and Sonar Systems Department} \\
\textit{Naval Undersea Warfare Center}\\
Newport, RI, USA \\
david.a.hague@ieee.org}
}
\maketitle

\begin{abstract}
The frequency dependent beampatterns of an active sonar projector filters the acoustic signal that is transmitted into the medium, also known as the transmit waveform. This filtering encodes information about the target's bearing relative to the main response axis. For any given projector and transmit waveform spectrum, there exists an optimal angle of operation which maximizes the Fisher Information (FI) of the target bearing estimate. Previous investigations into this phenomena show that for narrowband (i.e, high $Q$) Linear Frequency Modulated (LFM) waveforms, the angle of maximum FI is solely determined by its center frequency $f_c$. Steering the region of maximum bearing estimation precision is then achieved by appropriate selection of the LFM waveform's center frequency $f_c$. This fine bearing estimation is accomplished without steering the projector's main response axis.   In addition to LFM waveforms, a wide variety of other active sonar waveform types exist that possess distinct spectral characteristics.  These other waveforms possess different Ambiguity Function (AF) shapes from the LFM and are typically utilized to suite the range-Doppler resolution requirements of the active sonar system.  This paper investigates the transmit waveform impact on off-axis bearing estimation performance and the spectral filtering impact on the waveform's AF shape.  High $Q$ waveforms perform similarly to the LFM for off-axis bearing estimation while the transducer's spectral filtering perturbs the waveform's AF shape.
\end{abstract}


%=====================================================================================================================================================%
\section{Introduction}
Increased target localization precision is a primary goal of active sonar systems. Traditional systems increase localization precision by narrowing the beam of the projected acoustic energy to ensonify as small a region as possible. Beamwidth decreases as the ratio of aperture to the wavelength increases.  This results in increased localization precision but requires increasing the operating frequency and/or aperture of the system.  Increasing frequency will also increase propagation loss underwater while aperture is typically limited by the sonar platform's size.  Improving localization precision while mitigating the practical constraints of traditional narrow-beam systems is an open engineering challenge.

The authors of \cite{OffAxis}, \cite{biosonarBeam}, and \cite{Yang} investigated the behavior of echolocating mammals such as bats and dolphins which have exhibited angular resolution more precise than their respective beamwidths. Yovel \textit{et al.} observed Egyptian fruit bats alternating their sonar beam aim to either side of a landing platform rather than directly at it \cite{OffAxis}. They found that the observed beam angle relative to the platform maximized the gradient of broadband acoustic intensity as a function of angle.  This in turn maximizes the Fisher Information (FI) in a simple scalar signal model. The FI quantifies the curvature or sharpness of the log likelihood function for a given noise model \cite{KayEst}. This curvature determines the precision achievable by any unbiased estimator of the unknown parameter. The Cramer-Rao Lower Bound (CRLB) on the variance of any unbiased estimator is the inverse of the FI \cite{KayEst}. Arditi \textit{et al.} expanded on Yovel \textit{et al.}'s model to investigate how increasing the bat's mouth gape angle as well as the bandwidth of the transmitted signal affected this beam angle of maximum FI \cite{OffAxis}. Kloepper \textit{et al.} offered a more complete FI analysis including the impact of the signal spectrum, transducer response, and unknown echo amplitude \cite{Yang}. 

The aforementioned efforts \cite{OffAxis, biosonarBeam, Yang} took a waveform analysis approach by finding where the FI was maximized given a fixed transducer model and transmit waveform spectrum.  Recent work by Tidwell and Buck \cite{HeHeMatt} took a waveform synthesis approach where the FI was maximized given a fixed transducer model but varying transmit waveform spectrum.  They showed that for high $Q$ (i.e, narrowband) Linear Frequency Modulated (LFM) waveforms, its center frequency $f_c$ solely determined the angle at which the FI was maximized for the off-axis bearing estimation problem.  Ref \cite{HeHeMatt} utilized the LFM waveform, as many sonar systems do, due to its simple spectral structure and ease of implementation.  However, a wide variety of sonar waveform types exist \cite{Ricker} that possess distinct spectral characteristics.  These waveforms may be utilized because their Ambiguity Function (AF) suites the target time-delay (range) and Doppler (velocity) estimation precision requirements \cite{Ricker, Rihaczek} of the active sonar system.  This raises the intriguing question of whether the fine bearing estimates derived from \cite{Yang, HeHeMatt} may behave differently when transmitting a waveform with a more general spectral structure.  Additionally, it is likely that the sonar projector's frequency dependent filtering will also have an adverse impact on the waveform's AF shape degrading the waveform's ability to estimate and resolve target time-delay and Doppler.  

This paper takes a first step in addressing these questions by investigating both the transmit waveform impact on bearing estimation performance and the spectral filtering impact on the waveform's AF shape.  This research makes extensive use of the Multi-Tone Sinusoidal Frequency Modulated (MTSFM) waveform model, which is capable of synthesizing novel transmit waveforms with a broad array of AF shapes \cite{Hague_AES, Hague_JOE, Hague_EOA}.  High $Q$ MTSFM waveforms perform similarly to the LFM for off-axis bearing estimation with some carrier frequency offsetting which can be compensated for using the MTSFM waveform model.   The transducer's spectral filtering perturbs the waveform's AF shape, particularly the AF's sidelobe structure.  The rest of this paper is organized as follows: Section II gives an overview of the waveform signal model, off-axis bearing estimation, and the MTSFM waveform model.  Section III evaluates MTSFM waveforms using a similar method to that described in \cite{HeHeMatt} and compares off-axis bearing estimation performance as well as the impact of the transducer's spectral filtering on the waveform's AF shape.  Finally, Section IV concludes the paper.   

%=====================================================================================================================================================%

%=====================================================================================================================================================%
\section{Signal Model}

This section gives an overview of the waveform signal model, off-axis bearing estimation, and the MTSFM waveform model.  A bistatic active sonar system radiates acoustic energy into the medium via a directional electroacoustic transducer. The projected energy ensonifies a point scatterer and the return echo signal is measured using an omindirectional hydrophone. The range and velocity of the target are estimated from the time delay between the transmitted signal and received echo and the Doppler frequency shift of the received echo, respectively \cite{Ricker}.  The target's angular location is estimated based on the filtering the received echo undergoes due to the frequency dependent beampatterns of the projector.  

\subsection{The FM Waveform Model and the Ambiguity Function}
\label{subsubsec:model}
The FM waveform $s\left(t\right)$ is modeled as a complex analytic signal with unit energy and duration $T$ defined over the interval $-T/2 \leq t \leq T/2$.  The waveform is expressed in the time domain as
\begin{equation}
s\left(t\right) = \frac{\rect\left(\frac{t}{T}\right)}{\sqrt{T}}e^{j\varphi\left(t\right)}e^{2\pi f_c t}
\label{eq:ComplexExpo}
\end{equation}  
where $f_c$ is the waveform's carrier frequency, $\varphi\left(t\right)$ is the waveform's phase modulation function, and $\rect\left(\frac{t}{T}\right)$ is the rectangular amplitude tapering function normalized by the square root of the waveform's duration $T$ to ensure unit energy.  The waveform's instantaneous frequency is solely determined by its modulation function and is expressed as 
\begin{equation}
m\left(t\right) =  \dfrac{1}{2 \pi}\dfrac{\partial \varphi \left( t\right)}{\partial t}.
\label{eq:m}
\end{equation}  
This signal model assumes a Matched Filter (MF) receiver is used to process target echoes.  The MF, also known as a correlation receiver, is the optimal detection receiver for a known signal embedded in Additive White Gaussian Noise (AWGN) \cite{Rihaczek, Ricker}.  In a simple system where the receiver utilizes the transmit waveform as its MF, the MF will only be matched exactly to the target echo signal when that target is stationary relative to the system platform.  Targets with non-zero radial velocity known as range-rate $\dot{r}$ with respect to the system platform introduce a Doppler effect to the echo signal.  The general Doppler effect for broadband transmit waveforms compresses or expands the waveform in the time domain when the target is closing ($\dot{r}$ is positive) or receding ($\dot{r}$ is negative) respectively.  The Broadband Ambiguity Function (BAF) measures the response of the waveform's MF to its Doppler scaled versions and is defined as \cite{Ricker}
\begin{IEEEeqnarray}{rCl}
\chi\left(\tau, \eta\right) = \sqrt{\eta} \int_{-\infty}^{\infty}s\left(t\right)s^*\left(\eta \left(t+\tau \right) \right) dt
\label{eq:BAF}
\end{IEEEeqnarray}  
where $\tau$ represents time-delay and $\eta$ is the Doppler scaling factor expressed as
\begin{IEEEeqnarray}{rCl}
\eta = \left(\dfrac{1+\dot{r}/c}{1-\dot{r}/c}\right)
\label{eq:eta}
\end{IEEEeqnarray}  
where $c$ is the speed of propagation in the medium.  The BAF is the general model for analyzing the Doppler effect of broadband waveforms.  The quality factor $Q$ is a common measure of how broadband or narrowband a waveform is and is expressed as 
 \begin{IEEEeqnarray}{rCl}
Q = \dfrac{f_c}{\Delta f}
\label{eq:fracBand}
\end{IEEEeqnarray}
where $\Delta f$ is the waveform's swept bandwidth.  A higher $Q$ translates to an increasingly narrowband waveform.  

When $Q$ is considered large (i.e, $\geq 5$) and the ratio $\frac{\dot{r}}{c}$ is small, the Doppler effect is well approximated as a narrowband shift in the spectral content of the transmit waveform.  This is modeled by the Narrowband Ambiguity Function (NAF) which measures the response of the MF to the waveform's Doppler shifted versions and is defined as \cite{Rihaczek}
\begin{IEEEeqnarray}{rCl}
\chi\left(\tau, \nu\right) = \int_{-\infty}^{\infty}s\left(t\right)s^*\left(t+\tau\right)e^{j2\pi \nu t} dt
\label{eq:NAF}
\end{IEEEeqnarray}
where $\nu$ is the doppler shift expressed as $\nu = \frac{2\dot{r}}{c}f_c$.  While the BAF encompasses a more general model for the response of a waveform's MF to echos undergoing a Doppler effect, this paper focuses on narrowband transmit waveforms and thus the NAF is an appropriate measure of the waveform's MF output.  Going forward, this paper will refer to the NAF as simply the AF with the implication that the narrowband AF model is being employed.

Jointly estimating the time-delay and Doppler shift from a single target embedded in AWGN is performed by choosing the peak of the output of a bank of MF's tuned to different Doppler shifts.  The time-delay and Doppler corresponding to the peak MF output is taken as the target's time-delay and Doppler estimate.  For the case of a single target embedded in AWGN, the CRLB on the estimation variances for time-delay and Doppler are \cite{Ricker, Rihaczek} 
\begin{equation}
\text{var}\left(\hat{\tau}\right) \geq \left(\dfrac{1+SNR}{SNR^2}\right)\left(\dfrac{\tau_{rms}^2}{\beta_{rms}^2\tau_{rms}^2 - \gamma}\right)
\end{equation}
\begin{equation}
\text{var}\left(\hat{\nu}\right) \geq \left(\dfrac{1+SNR}{SNR^2}\right)\left(\dfrac{\beta_{rms}^2}{\beta_{rms}^2\tau_{rms}^2 - \gamma}\right)
\end{equation}
where SNR is the signal to noise ratio at the output of the MF, $\tau_{rms}^2$ is the waveform's Root Mean Square (RMS) pulse-length, $\beta_{rms}^2$ is the waveform's RMS bandwidth, and $\gamma$ is the Range-Doppler Coupling Factor (RDCF).  The RMS bandwidth, pulse-length, and RDCF are expressed respectively as \cite{Ricker, Rihaczek} 
\begin{align}
\beta_{rms}^2 &= 4\pi^2\int_{-\infty}^{\infty}\left(f - f_0\right)^2 |S\left(f\right)|^2 df,  \\
\tau_{rms}^2  &= 4\pi^2\int_{\Omega_t} \left(t-t_0\right)^2 |s\left(t\right)|^2 dt, \\
\gamma &= -4\pi^2 \mathbb{I} \Biggl\{\int_{\Omega_t} tm\left(t\right) dt \Biggr\}
\label{eq:EOA}
\end{align}
where $f_0$ is the waveform's spectral centroid $\langle f \rangle$, $t_0$ is the first time moment $\langle t \rangle$ of the the waveform, $S\left(f\right)$ is the waveform's Fourier transform, and $\Omega_t$ represents the region of support in time of the waveform.  These three parameters are a function of the envelope and modulation function characteristics of the waveform \cite{Rihaczek}.  Designing a waveform with $\gamma = 0$ will minimize the variance on the joint time-delay and Doppler estimates. 

\subsection{Off-Axis Bearing Estimation}
\label{subsec:bearEst}
This paper focuses on estimating the target's angular location via the filtering the received echo undergoes from the frequency dependent beampatterns of the projector transducer.   The beamwidth of the transducer's Main Response Axis (MRA) generally decreases as frequency increases and is equivalent to an angle dependent spectral filter.  This filter's frequency response encodes information about the direction of the signal relative to the MRA. Arditi \textit{et al.} derived the maximum likelihood estimator for the unknown target bearing in \cite{biosonarBeam} as the bearing whose corresponding expected received signal produces the largest normalized correlation with the actual received signal.

The received signals are described by a linear model with range of frequencies $f_1,...,f_N$
\begin{equation}
\label{eq:model}
	\begin{aligned}
    \begin{bmatrix}
        y_1(\theta) \\
        y_2(\theta) \\
        \vdots \\
        y_N(\theta) \\
    \end{bmatrix} &= a
    \begin{bmatrix}
        S_1b(f_1,\theta) \\
        S_2b(f_2,\theta) \\
        \vdots \\
        S_Nb(f_N,\theta) \\
    \end{bmatrix} +
    \begin{bmatrix}
        w_1 \\
        w_2 \\
        \vdots \\
        w_N \\
    \end{bmatrix}
    \\
    \boldsymbol{y} &= a \ \boldsymbol{h} + \boldsymbol{w} \ ,
	\end{aligned}
\end{equation}
where $\theta$ is the target bearing relative to the beam's main axis, $y_i(\theta)$ is the received signal at frequency $f_i$, $a$ is the unknown amplitude of the received signal echo due to both geometric spreading and unknown target strength, which is assumed to be independent of frequency.  The quantity $S_i = S(f_i)$ is the transmitted signal at frequency $f_i$, $b(f_i,\theta)$ is the beampattern evaluated at frequency $f_i$ and bearing $\theta$, and $w_i$ is statistically independent AWGN with equal power ($\sigma^2$) across all frequencies \cite{Yang}.

The CRLB on the variance of the bearing estimate inferred from this model is a natural metric for the performance of the system. Reference \cite{Yang} defines the CRLB on the bearing estimate as:
\begin{equation}
\label{eq:FI}
    \textrm{var}(\hat{\theta}) \geq
    \begin{bmatrix}
        \boldsymbol{J}^{-1}(\theta,a) 
    \end{bmatrix} _{11} = \biggl(\textrm{SNR} \  \norm{ \frac{\partial \boldsymbol{h}}{\partial \theta} } ^2 \ \sin^2(\psi)\biggr)^{-1} ,
\end{equation}
where \textbf{J} is the Fisher Information matrix, SNR = $(a/\sigma)^2$, $\norm{\cdot}$ is the Euclidean norm, and $\psi$ is the principal angle between $\boldsymbol{h}$ and $\partial \boldsymbol{h} / \partial \theta$ \cite{golub1996matrix}. The reciprocal of (\ref{eq:FI}) is referred to as bearing FI for convenience.

\subsection{Multi-Tone Sinusoidal Frequency Modulation}
\label{subsec:MTSFM}

The MTSFM waveform is synthesized by representing the modulation function \eqref{eq:m} as a Fourier series expansion. The modulation function is expressed in terms of even and odd symmetric harmonics as
\begin{align}
m\left(t\right) &= m_e\left(t\right) + m_o\left(t\right) \\ &= \frac{a_0}{2} + \sum_{k=1}^K a_k \cos\left(\frac{2 \pi k t}{T}\right) + b_k \sin\left(\frac{2 \pi k t}{T}\right).
\label{eq:MTSFM_1}
\end{align}
where $m_e\left(t\right)$ and $m_o\left(t\right)$ are respectively the even and odd symmetric components of the Fourier series expansion
\begin{align}
m_e\left(t\right) &= \frac{a_0}{2} + \sum_{k=1}^Ka_k \cos\left(\frac{2 \pi k t}{T}\right),\label{eq:memo1} \\
m_o\left(t\right)&= \sum_{k=1}^Kb_k \sin\left(\frac{2 \pi k t}{T}\right). 
\label{eq:memo2}
\end{align}
Integrating with respect to time and multiplying by $2\pi$ yields the phase modulation function of the waveform expressed as
\begin{align}
\varphi\left(t\right) &= \varphi_e\left(t\right)+ \varphi_o\left(t\right) \\ &= \pi a_0t + \sum_{k=1}^K \alpha_k \sin\left(\frac{2 \pi k t}{T}\right) - \beta_k \cos\left(\frac{2 \pi k t}{T}\right)
\label{eq:MTSFM_2}
\end{align}
where $\alpha_k$ and $\beta_k$ are the MTSFM waveform's modulation indices and $\varphi_e\left(t\right)$ and $\varphi_o\left(t\right)$ are the instantaneous phase functions derived from the even and odd modulation functions \eqref{eq:memo1} and \eqref{eq:memo2}
\begin{align}
\varphi_e\left(t\right) &= \pi a_0t + \sum_{k=1}^K \alpha_k \sin\left(\frac{2 \pi k t}{T}\right), \\
\varphi_o\left(t\right) &= -\sum_{k=1}^K \beta_k \sin\left(\frac{2 \pi k t}{T}\right) 
\label{eq:MTSFM_3}
\end{align}
The even/odd modulation functions are explicitly defined here because MTSFM waveforms with either even or odd symmetry in their modulation functions have distinct spectral characteristics which will be discussed later in the paper.  Inserting \eqref{eq:MTSFM_2} into the waveform signal model \eqref{eq:ComplexExpo} yields the MTSFM waveform time-domain representation
\begin{multline}
s\left(t\right) = \dfrac{\rect\left(t/T\right)}{\sqrt{T}} \times \\ \exp\Biggl\{j\sum_{k=1}^K \alpha_k \sin\left(\frac{2 \pi k t}{T}\right) - \beta_k \cos\left(\frac{2 \pi k t}{T}\right) \Biggr\}.
\label{eq:MTSFM_4}
\end{multline}
The modulation indices $\alpha_k$ and $\beta_k$ act as a discrete set of parameters that may be adjusted to synthesize novel waveform designs.  The MTSFM waveform was first used in \cite{Hague_JOE} to represent the modulation function of a family of Doppler sensitive waveforms.  More recently, efforts in \cite{Hague_AES} and \cite{Hague_EOA} have defined optimization methods to design MTSFM waveforms with very specific and unique characteristics.  

The MTSFM waveform in \eqref{eq:MTSFM_4} can also be defined as a complex Fourier series expansion
\begin{IEEEeqnarray}{rCl}
 s\left(t\right) =  \dfrac{\rect\left(t/T\right)}{\sqrt{T}}\sum_{m=-\infty}^{\infty} c_m  e^{j\frac{2\pi m t}{T}}e^{j\pi a_0 t}.
\label{eq:MTSFM_5}
\end{IEEEeqnarray}
The complex Fourier coefficients are expressed in exact closed form in terms of three types of Generalized Bessel Functions (GBF) depending on the symmetry of the waveform's modulation function \cite{Hague_AES}
\begin{equation}  c_m = \left\{
\begin{array}{ll}
	\mathcal{J}_m^{1:K}\left(\{\alpha_k, -j\beta_k\}\right), & \varphi\left(t\right) \\

      \mathcal{J}_m^{1:K}\left(\{\alpha_k\}\right), & \varphi_e\left(t\right) \\
      
      \mathcal{J}_m^{1:K}\left(\{-\beta_k\}, \{-j^k\}\right), & \varphi_o\left(t\right) \\
\end{array} 
\right.
\label{eq:SFM_Fourier_Series} 
\end{equation}
where $\mathcal{J}_m^{1:K}\left(\{\alpha_k, -j\beta_k\}\right)$ is the $K$-dimensional GBF of the mixed-type, $\mathcal{J}_m^{1:K}\left(\{\alpha_k\}\right)$ is the cylindrical $K$-dimensional GBF, and $\mathcal{I}_m^{1:K}\left(\{-j\beta_k\}\right)$ is the $K$-dimensional Modified GBF (M-GBF) \cite{Lorenzutta}.  The various forms of GBFs are K-dimensional generalizations of the standard 1-dimensional Bessel functions and share many of their fundamental properties \cite{Dattoli}.  This representation of the MTSFM readily admits a closed-form expression for the MTSFM's spectrum which is expressed as \cite{Hague_AES}
\begin{multline}
S\left(f\right) = \sqrt{T} \times \\ \sum_{m=-\infty}^{\infty}\mathcal{J}_m^{1:K}\left(\{\alpha_k,-j\beta_k\}\right)\sinc\left[\pi T \left(f-\frac{m}{T}\right)\right].
\label{eq:MTSFM_Spec}
\end{multline}
The complex Fourier series representation of the MTSFM waveform in \eqref{eq:SFM_Fourier_Series} and \eqref{eq:MTSFM_Spec} also readily allows for deriving closed-form expressions for its AF \cite{Hague_AES}.  The MTSFM waveform naturally possesses a constant envelope and the vast majority of its energy will be densely concentrated in its swept bandwidth $\Delta f$ \cite{Hague_AES}.  These two properties are necessary for efficient transmission on practical piezo-electric transducers \cite{Hague_JASA}.

\begin{figure}[hthp]
\centerline{\includegraphics[width = 0.5\textwidth]{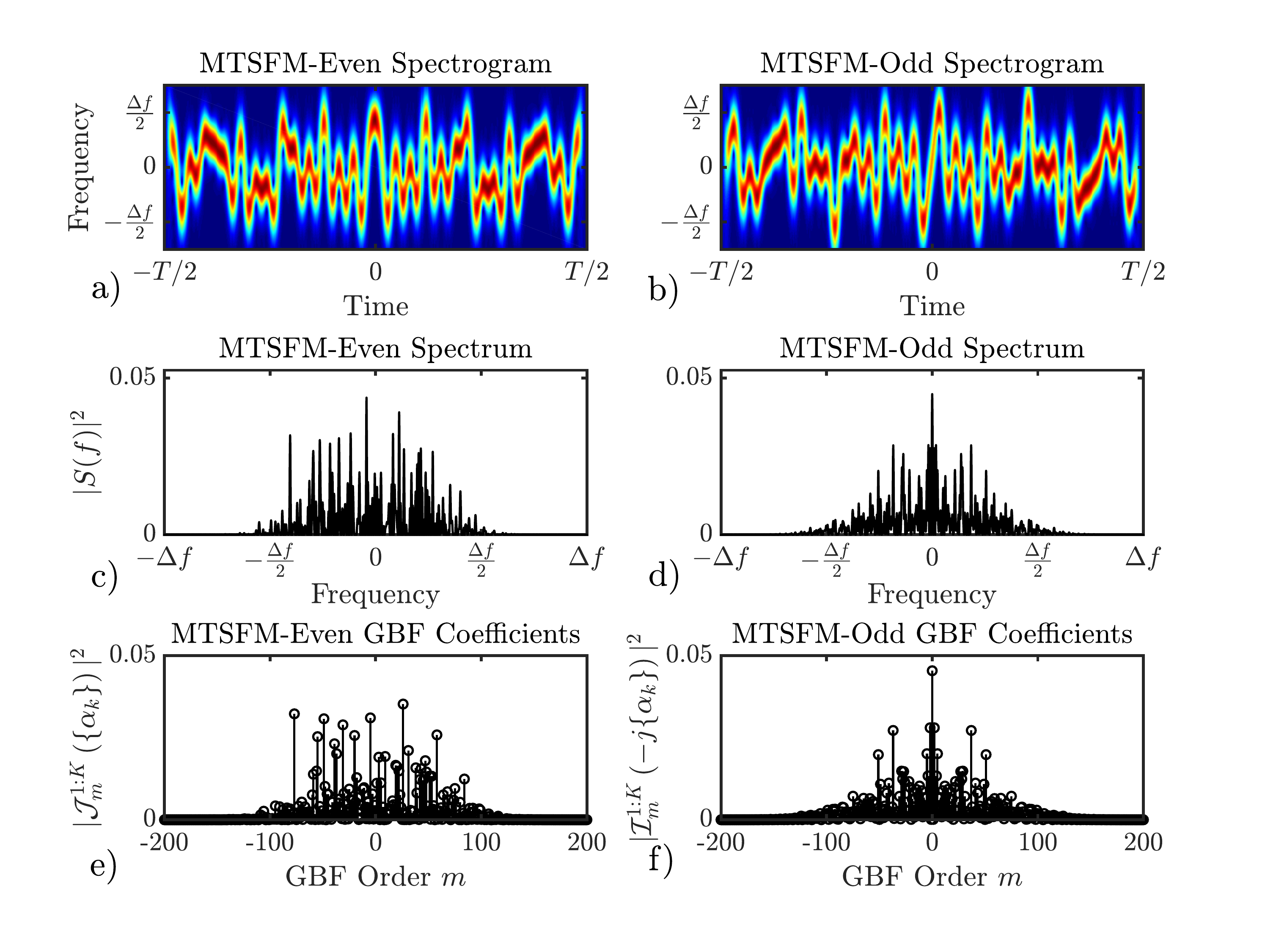}}
\caption{Spectrograms (a)-(b), spectra (c)-(d), and GBF coefficients (e)-(f) of MTSFM waveforms employing even and odd symmetry in their modulation functions.  The MTSFM with an even-symmetric modulation function possesses a non-symmetric spectrum while the MTSFM with an odd-symmetric modulation function possesses even symmetry in its spectrum.  The spectral shapes are formed from the GBF coefficients which also possess the same symmetry properties as the corresponding waveform's spectra.}
\label{fig:MTSFM_Spectra}
\end{figure}

Figure \ref{fig:MTSFM_Spectra} shows the spectrogram, spectrum, and GBF coefficients for MTSFM waveforms employing even and odd symmetry in their modulation functions.  Both waveforms utilize the same $K=32$ coefficients used in the illustrative design example shown in \cite{Hague_AES}.  However, each waveform utilizes a different set of harmonics in their respective modulation functions.  The first waveform (left panels of the figure) possesses a modulation function composed soley of cosine harmonics and exhibits even-symmetry in time.  The second waveform (right panels of the figure) possesses a modulation function composed of sine harmonics soley and exhibits odd-symmetry in time.  The symmetry of the modulation function has a profound impact on the symmetry of the resulting waveform's spectrum.  The first waveform's even-symmetric modulation function results in a non-symmetric spectrum whereas the second waveform's odd-symmetric modulation function results in an even-symmetric spectrum.  These spectral characteristics of these waveforms are directly related to the symmetry of the GBF coefficients that represent each waveform's spectrum.  The symmetry in the waveform's modulation function also has a profound impact of the waveform's ability to jointly estimate a target's time-delay and Doppler and can be directly analyzed via $\gamma$ in \eqref{eq:EOA}.  Any waveform with an even-symmetric modulation function will possess a RDCF $\gamma$ of exactly zero \cite{Ricker}.  Waveforms with odd-symmetry in their modulation functions can possess a RDCF that ranges from almost zero to, in the case of an LFM, perfectly coupled measurements \cite{Rihaczek}.  As will be shown in the next section, the unique spectral symmetry properties of the MTSFM also have an impact on off-axis bearing estimation performance.

%=====================================================================================================================================================%

%=====================================================================================================================================================%
\section{Joint Evalutation of Off-Axis Bearing Estimation Performance and Waveform Range-Doppler Characteristics}
\label{sec:Simulations}

\subsection{Methods}
\label{subsec:Methods}
This work utilizes the same continuous line source (CLS) from \cite{HeHeMatt} with length $L$ as a projector transducer with beampattern
\begin{equation}
	\label{eq:bp}
	b(f_i,\theta) = \frac{\sin(\frac{1}{2}k_iL\sin(\theta))}{\frac{1}{2}k_iL\sin(\theta)} \ ,
\end{equation}
where $k_i = 2\pi f_i/c$ and $c$ = 1500 m/s is the speed of acoustic propagation in water \cite{Kinsler}. The transmitter generates MTSFM waveforms with a spectrum as defined in \eqref{eq:MTSFM_Spec} which is a function of the modulation indicies $\alpha_k$ and $\beta_k$.  As was shown in \cite{HeHeMatt}, for narrowband (i.e, high $Q$) waveforms, the angle $\Theta^*$ that maximizes the FI is solely determined by the waveform's center frequency $f_c$.  Thus, changing $f_c$ will result in maximum bearing precision at different angles of arrival without physically or electronically steering the transducer's main response axis.  The challenge now is to determine if this same technique can also be applied to high $Q$ MTSFM waveforms which possess novel modulation functions and more general spectral characteristics.  Additionally, the same spectral filtering that allows for this fine bearing precision also filters the waveform echo itself.  Since these waveforms sweep through a bandwidth $\Delta f$ in a periodic fashion, the spectral filtering will amplify and attenuate the waveform at different frequencies introducing oscillatory Amplitude Modulation (AM) effects to the waveform time series.  This AM effect will likely perturb the waveform's AF shape.  Therefore, the following simulations will not only evaluate the off-axis bearing performance of the MTSFM waveforms, but also evaluates the impact of the spectral filtering on the waveform's AF mainlobe and sidelobe characteristics. 

\subsection{Off-Axis Bearing Estimation Performance}
\label{subsec:Off_Axis_Perf}

\begin{figure}[!ht]
\centering
\includegraphics[width=0.5\textwidth]{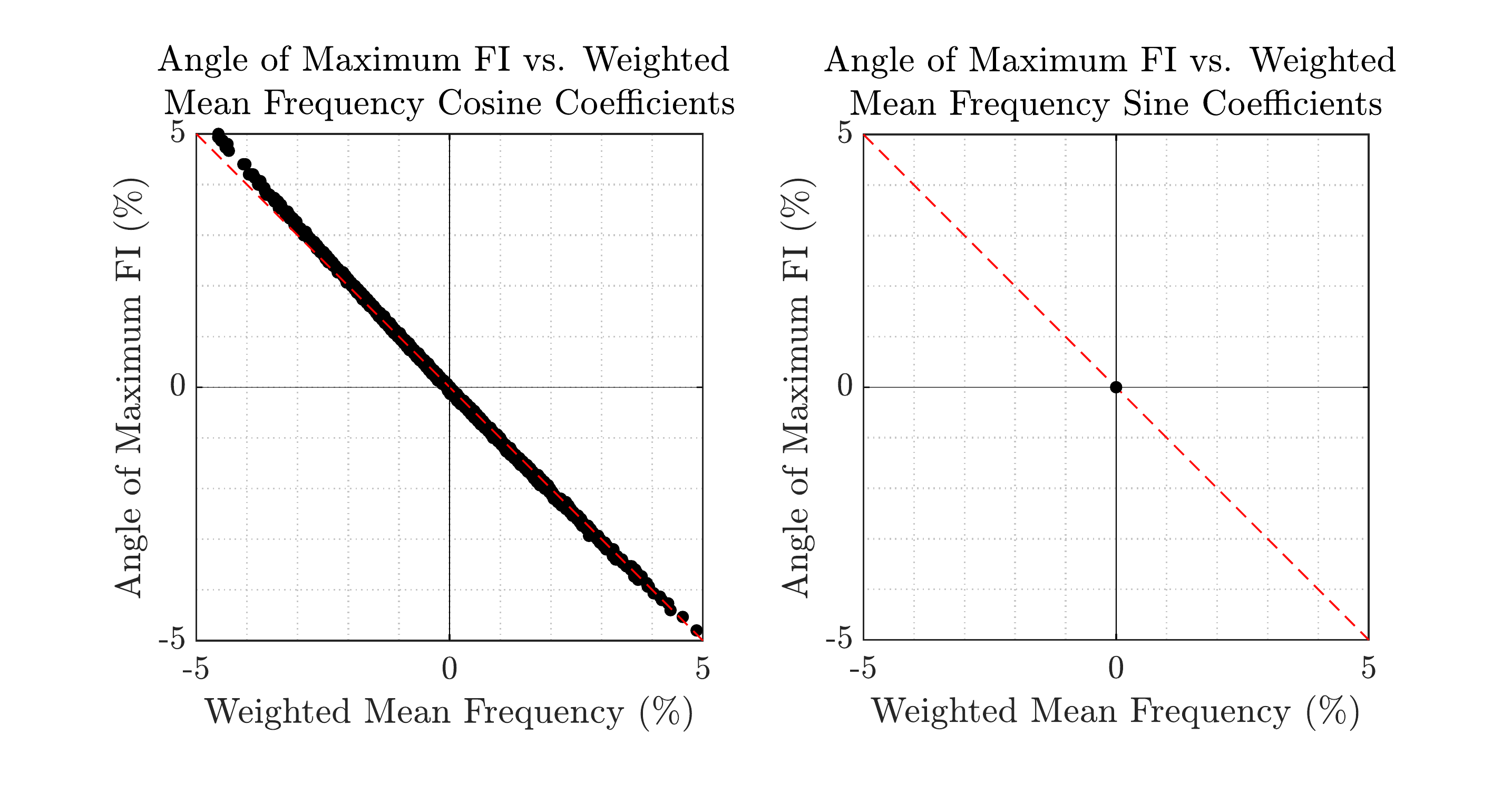}
\caption{Angle of Maximum FI vs. weighted mean frequency for two sets of 1000 MTSFMs whose modulation functions are composed either solely of cosine or sine coefficients.  The MTSFMs with cosine coefficients possesses a non-symmetric spectrum resulting in a deviation in mean frequency and therefore deviation in the angle of maximum FI.  The MTSFMs with sine coefficients possess an even-symmetric spectrum like the LFM resulting in zero deviation in mean frequency and zero deviation in the angle of maximum FI.}
\label{fig:MTSFM_2}
\end{figure}

Figure \ref{fig:MTSFM_2} shows percent deviation in the angle of the maximum FI $\Theta^*$ as a function of the percent deviation of weighted mean frequency $f_c$ for two sets of 1000 MTSFMs composed of cosine and sine harmonics in their respective modulation functions.  Each waveform sweeps through a band of frequencies $f_c \pm \Delta f/2$ and possesses a Time-Bandwidth Product (TBP) $T \Delta f = 100$ and $Q=5$.  The MTSFM with cosine coefficients possesses even-symmetry in its modulation function.  As a result of this, the spectral shape of the resulting MTSFM waveform is not even-symmetric resulting in a shift $\delta f$ of the weighted mean frequency of the waveform.  This deviation in mean frequency is generally much less than the waveform's swept bandwidth $\Delta f$ and is nearly perfectly correlated with the deviation in angle of maximum FI.  The MTSFM with sine coefficients possesses odd-symmetry in its modulation function resulting in an even-symmetric spectrum much like the LFM waveforms evaluated in \cite{HeHeMatt} and there is no measureable deviation in the angle of maximum FI.    

\begin{figure}[hthp]
\centerline{\includegraphics[width = 0.5\textwidth]{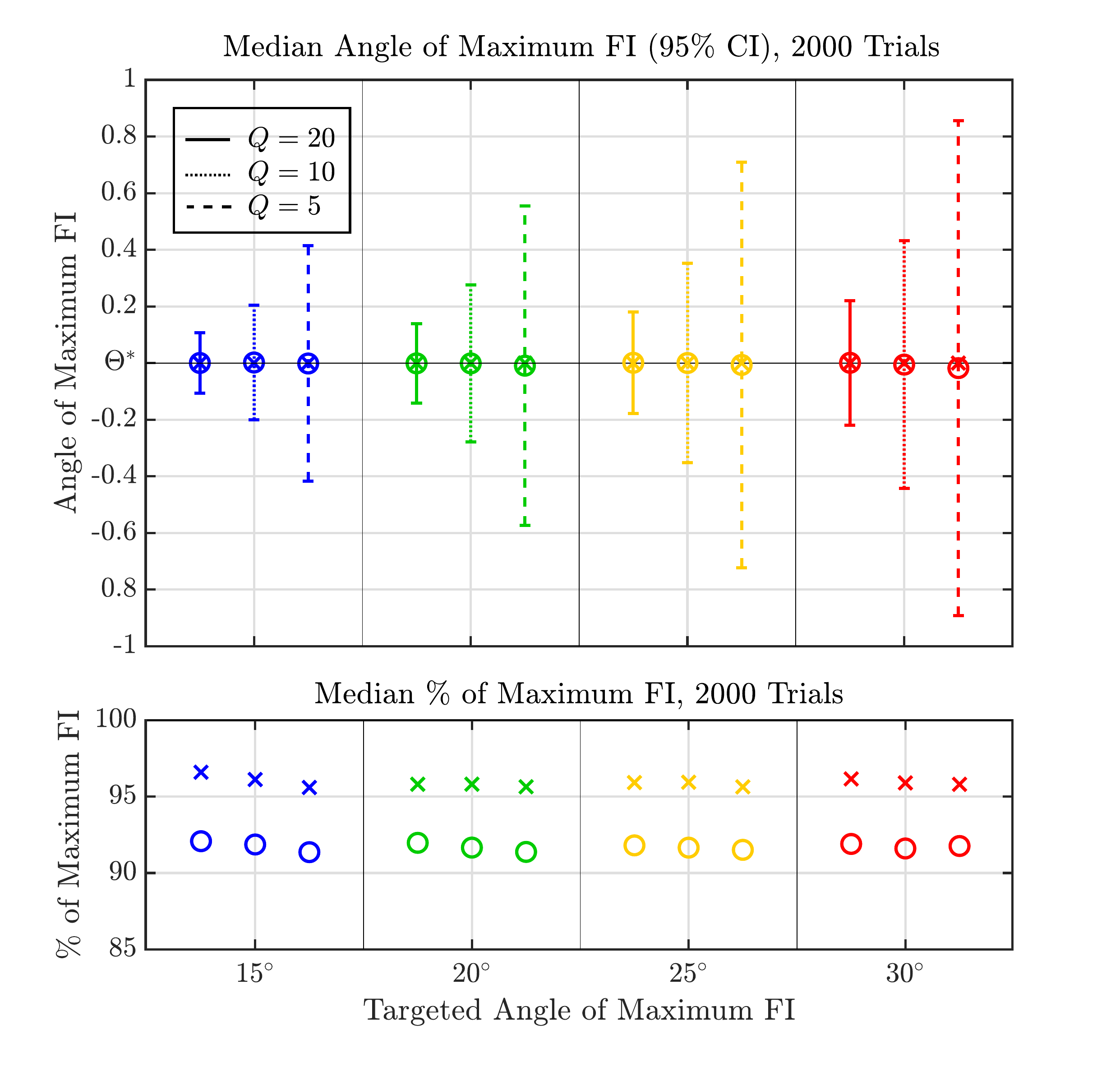}}
\caption{Top Panel : Median (denoted by x's), mean (denoted by circles), and error bars representing the $95\%$ confidence intervals of the angle of maximum FI for 2000 randomly generated MTSFM waveforms with even symmetric modulation functions for fixed $f_c$ but varying $Q$ (i.e, increasing bandwidth) for a range of specific maximum FI angles.  Bottom Panel : mean and median of the same waveform trials as a percentage of the maximum FI value.  The off-axis angle of arrival estimation deviation increases with decreasing $Q$ but does not substantially alter the maximum FI value.}
\label{fig:MTSFM_3}
\end{figure}

Further inspection of the MTSFM waveforms with even-symmetric modulation functions shows this deviation is exacerbated with increasing bandwidth across a variety of desired angles of maximum FI $\Theta^*$.  Figure \ref{fig:MTSFM_3} shows error bars which plot the deviation in $\Theta^*$ for MTSFM waveforms with decreasing $Q$.  The x's denote the median and the circles denote the mean of the trials.  The bottom panel shows the median and mean $\%$ of maximum FI for each of the targeted angles of maximum FI.  For each targeted $Q$ value, 2000 MTSFM waveforms were generated each possessing the same $f_c$ but increasing bandwidth $\Delta f$.  Waveforms with $Q = 20,~10,~\&~5$ have corresponding TBPs of 100, 200, and 400 respectively.  The bars represent the $95\%$ confidence intervals of the trials.  Decreasing $Q$ corresponded on average to an increase in deviation in the off-axis bearing estimate.  While the deviation in bearing estimates increase with decreasing $Q$, the values of the FI function stayed nearly the same.  This is an expected result as \cite{HeHeMatt} showed that the FI function possesses a distinct peak value whose width increases with decreasing $Q$.  The increased peak width would facilitate greater deviation in bearing estimates while still nearly attaining the maximum FI value.  One way to mitigate this deviation in the bearing estimate is to calculate the deviation in frequency $\delta f$ from $f_c$ resulting from the MTSFM waveform's spectrum and offset it by setting $a_0 = -2\delta f$ in \eqref{eq:MTSFM_1}.  The MTSFM will now sweep through a band of frequencies $f_c + a_0/2 \pm \Delta f/2$ and possess a mean frequency of $f_c$ thus removing the deviation in the angle of maximum FI.

\subsection{MTSFM Waveform Range-Doppler Characteristics}
\label{subsec:MTSFM_RangeDop}

\begin{figure}[!h]
\centering
\includegraphics[width=0.5\textwidth]{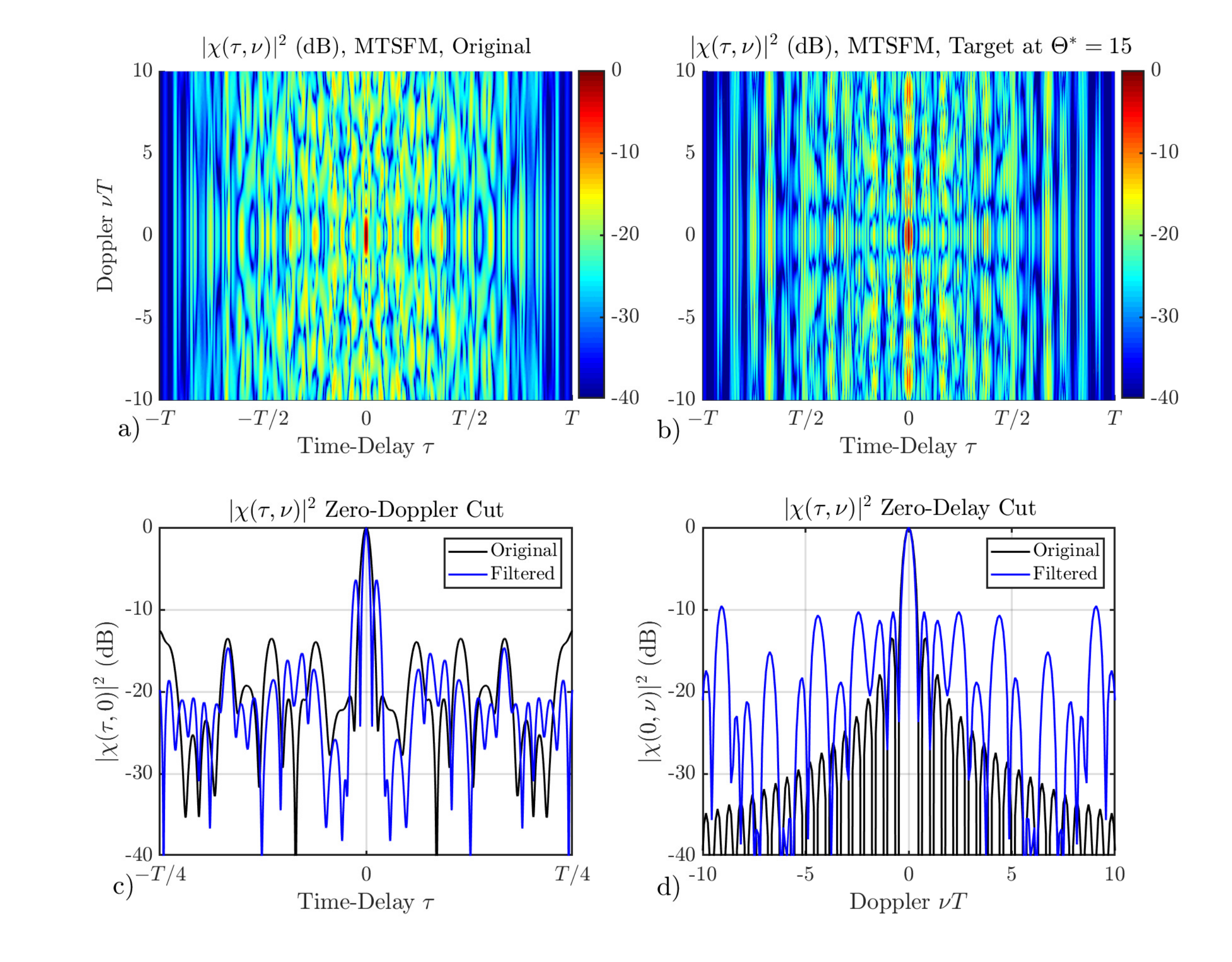}
\caption{AFs (a)-(b), zero-Doppler cut (c), and zero-time-delay cut (d) of MTSFM with a TBP of 100 and $Q = f_c/\Delta f=5$ with and without the transducer's spectral filtering.  The transducer's spectral filtering degrades the mainlobe and sidelobe structure of the waveform's AF shape.}
\label{fig:MTSFM_AF}
\end{figure}

Figure \ref{fig:MTSFM_AF} shows the impact of the transducer's spectral filtering on the AF of a MTSFM waveform with cosine coefficients.  Also shown in in Figure \ref{fig:MTSFM_AF} are the zero time-delay and zero-Doppler cuts of the AF.  The spectral filtering introduces AM perturbations to the waveform time-series.  These effects combine to produce destructive interference in the AF's mainlobe and sidelobe structure, particularly in the AF's zero-Doppler cut.   These perturbations are not severe enough to substantially degrade the AF's zero-time-delay mainlobe width, but the sidelobe levels are notably increased, particularly so in Doppler. 

%=====================================================================================================================================================%

%=====================================================================================================================================================%

%=====================================================================================================================================================%
\section{Conclusions}
The MTSFM waveform designs presented in this paper maximize the bearing Fisher Information for the active sonar localization problem described in \cite{Yang} at a desired bearing using a continuous line source transducer.  As with the previous efforts of \cite{HeHeMatt}, the region of maximum bearing estimation precision is achieved by changing the center frequency $f_c$ of these high $Q$ MTSFM waveforms without physically or electronically steering the transducer's MRA.  Simulations show that the MTSFM waveforms with odd-symmetry in their modulation functions perform essentially the same as the LFM waveforms with equivalent $Q$ and center frequency $f_c$ due to their even-symmetric spectral shapes.  MTSFM waveforms with even-symmetry in their modulation functions have non-symmetric spectra resulting in a shift in center frequency $f_c$ which biases the angle of arrival estimate.  This bias can be eliminated by modifying the waveform's swept band of frequencies.  The spectral filtering from the transducer itself introduces perturbations to the waveform in the form of AM effects which degrades the mainlobe and sidelobe structure of the waveform's AF.  Future efforts will focus on evaluating this off-axis bearing estimation method for broadband (i.e, low $Q$) waveforms and for other transducer beampatterns. 

%=====================================================================================================================================================%
%\bibliographystyle{IEEEtran}
%\bibliography{Tidwell_Hague_OCEANS_2021}

%=====================================================================================================================================================%

%=====================================================================================================================================================%

%
\end{document}